\documentclass[11pt]{article}
\usepackage{graphicx}
\usepackage[margin=1.25in]{geometry}
\usepackage[usenames,dvipsnames]{color}
\usepackage{url}
\usepackage{hyperref}
\hypersetup{colorlinks = true,
            linkcolor = blue,
            urlcolor  = blue,
            citecolor = blue,
            anchorcolor = blue}
\usepackage{siunitx}
\usepackage{tabulary}
\usepackage{blindtext}
\usepackage{graphicx}
\graphicspath{{Figures/}}
\usepackage{siunitx}
\usepackage{chemformula}
\usepackage{hyperref}

\def\Title#1{\begin{center} {\LARGE #1 } \end{center}}
\def\Author#1{\begin{center}{ \sc #1} \end{center}}
\def\Address#1{\begin{center}{ \it #1} \end{center}}

\newenvironment{Abstract}{\begin{quotation} \begin{center}
                       ABSTRACT
     \end{center}\bigskip  }{\end{quotation}}

\newcommand\snowmass{\begin{center}\rule[-0.2in]{\hsize}{0.01in}\\\rule{\hsize}{0.01in}\\
\vskip 0.1in Submitted to the  Proceedings of the US Community Study\\ 
on the Future of Particle Physics (Snowmass 2021)\\ 
\rule{\hsize}{0.01in}\\\rule[+0.2in]{\hsize}{0.01in} \end{center}}

\begin{document}


\Title{Plasma Processing for In-Situ Field Emission Mitigation of Superconducting Radiofrequency (SRF) Cryomodules}


\Author{M. Martinello, P. Berrutti, B. Giaccone, S. Belomestnykh, M. Checchin, G.V. Eremeev, A. Grassellino, T. Khabibouilline, A. Netepenko, R. Pilipenko, A. Romanenko, S. Posen, G. Wu}

\Address{%
Fermi National Accelerator Laboratory, Kirk Rd and Pine St, Batavia, IL 60510, USA
}%

\Author{D. Gonnella, M. Ross, J.T. Maniscalco}

\Address{%
SLAC National Accelerator Laboratory, 2575 Sand Hill Rd, Menlo Park, CA 94025, USA
}%

\Author{T. Powers}

\Address{%
Thomas Jefferson National Accelerator Facility, Newport News, VA 23606, USA
}%

\medskip

 \begin{Abstract}
\noindent Field emission (FE) is one of the main limiting factors of superconducting radio-frequency (SRF) cavities operating in accelerators and it occurs whenever contaminants, like dust, metal flakes or even absorbates, are present on the surface of the cavity high electric field region. Field emission reduces the maximum achievable accelerating field and generates free electrons that may interact with the beam, damage or activate the beamline. One practical method that can be used to mitigate this problem is in-situ plasma cleaning, or plasma processing. The development of a processing that can be applied in-situ is extremely advantageous, since it enables the recovery of the cryomodule performance without the need of disassembling the whole cryomodule, which is an extremely expensive and time-consuming process. On the other hand, plasma processing only requires the cryomodule warm-up to room-temperature and the subsequent processing of the contaminated cavities. The entire process is reasonably quick and involves a limited number of personnel. For these reasons we would like to advocate for continuing to invest in the R\&D of plasma processing to optimize its applicability in cryomodules and for extending the technique to other frequency ranges and cavities geometries.

\end{Abstract}

\snowmass

\def\thefootnote{\fnsymbol{footnote}}
\setcounter{footnote}{0}

\section{Introduction}

Superconducting radio-frequency (SRF) cavities are key components of modern particle accelerators, and the continuous improvement of their efficiency is critical to realize affordable and more powerful particle accelerators which are needed to carry out pivotal high energy physics experiments. 
State-of-the-art SRF cavities are very efficient devices that can achieve accelerating gradients up to about 50 MV/m during vertical RF test. However, when assembled together into a string of cavities – the main component of a cryomodule – there are high chances of introducing contaminants which will field emit as the electromagnetic field is established in the resonator. 
The emitted electrons are accelerated by the electric field and impact the cavity walls depositing heat and creating bremsstrahlung x-rays and neutrons. The x rays produced by the FE can cause radiation damage to the cryomodule’s components, decreasing operational lifetime. Once the FE is activated, it limits the cavity’s accelerating field and causes a degradation in quality factor due to the additional dissipation introduced by the emitted electrons. If FE is severe, it can cause thermal breakdown of the cavity and can also activate the beam line, causing induced radioactivity in the cavity. Sources of FE are contaminants (dust or metal particles) or cavity surface defects that cause local enhancement of the FE current. Field emitters may also turn on after the cavities are installed in the tunnel due to particle migration in the beam line. In addition, the presence of even a few monolayers of hydrocarbons, or other adsorbate gases, on the cavity surface can further decrease the Nb work function \cite{bagus2008work}, facilitating FE.  The origin of hydrocarbon contamination on the cavity inner surface is not completely understood; however, its presence has been reported in the literature in multiple cases \cite{doleans2016situ, cao2013} and carbon has been observed in both adventitious form and as local contamination on the Nb surface. Doleans \textit{et al.} report in \cite{doleans2016situ} that evidence of volatile hydrocarbon has been found through residual gas analysis on thermally cycled SNS (Spallation Neutron Source) cryomodules; they explain that these signals must originate from the released gases that were previously condensed on the cavity walls at cryogenic temperature or from species produced during accelerator operation by the interaction of electrons with the cavity surface contaminants.
Plasma cleaning can be used on SRF cavities in-situ in the cryomodules to remove the hydrocarbon contamination and restore the niobium work function obtaining a decrease in FE and a corresponding increase in the accelerating gradient. 
Plasma cleaning consists in a process in which impurities are removed from a surface via a mixture of inert and reactive gases in the form of a glow discharge. Plasma processing was developed for SRF cavities some years ago by Oak Ridge National Laboratory (ORNL). They demonstrated that a plasma composed by a mixture of neon and a small percentage of oxygen reduced hydrocarbon-related field emission in the Spallation Neutron Source (SNS) cavities \cite{doleans2013plasma, doleans2016ignition, doleans2016plasma, doleans2016situ}. Starting from this successful experience, plasma processing studies are being conducted at multiple laboratories for different accelerating structures \cite{legg_plasma, giaccone2021splasma, berrutti2019plasma, giaccone2021field, powersplasma, zhangplasma, wu2018situ, wu2019cryostat, huang2019effect}.

\section{Plasma processing effort at Fermilab}

\subsection{Plasma cleaning of SRF cavities }

At Fermilab, the majority of the plasma processing effort focused on developing plasma cleaning for LCLS-II and LCLS-II HE cavities and cryomodules, in collaboration with ORNL and SLAC National Accelerator Laboratory.

\begin{figure}[h]
    \centering
    {\includegraphics[width=1\columnwidth]{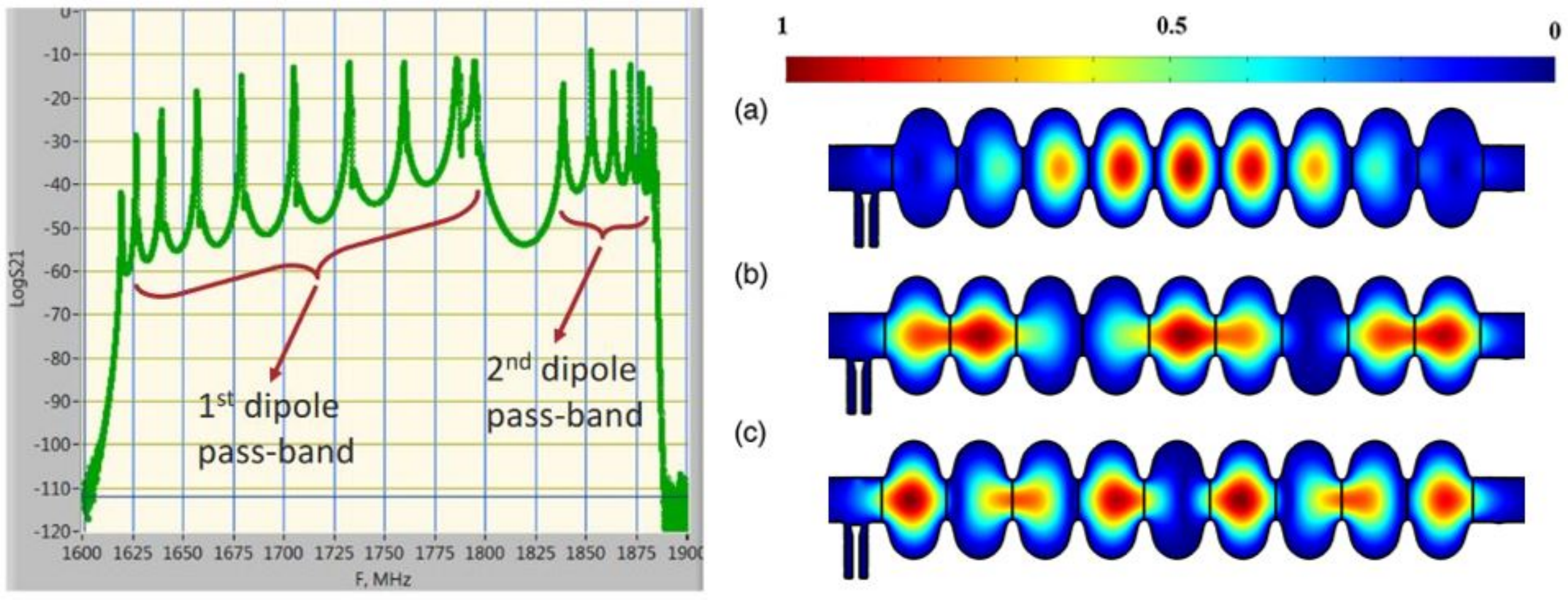}} \quad
	\caption{S21 measurement of the first and second dipole pass band of a 1.3 GHz 9-cell cavity (left); normalized electric field amplitude along the X-Z plane of the modes used for igniting the plasma in cell 5 and for transferring the plasma from cell 5 to cell 6: mode 2-1 (a), mode 1-3 (b), mode 1-6 (c) (right).}
	\label{fig:HOMign}
\end{figure}

A new method of plasma ignition for LCLS-II 1.3 GHz TESLA-shaped cavities was developed by Berrutti \cite{berrutti2019plasma}.
The method consists of superimposing two HOMs (of the first and second dipole pass band) to ignite the glow discharge in one of the cavity cells, and to move it to the other cells. Figure \ref{fig:HOMign} shows on the left the S21 measurament of the first and second dipole pass band, and on the right the electric field amplitude along the X-Z plane of the modes used for igniting the plasma in cell 5 -- first mode of the second dipole band (mode 2-1) (a) -- and for transferring the plasma from cell 5 to cell 6 -- third mode of the first dipole band (mode 1-3) (b) and sixth mode of the first dipole band (mode 1-6) (c). The complete sequence is the following: after ignition of cell 5 with the mode 2-1, the plasma can be locked into the same cell superimposing mode 1-5 and then turning off mode 2-1; the plasma can then be transferred to cell 6 superimposing mode 1-3; then mode 1-5 can be shut off and mode 1-6 can be used to lock the plasma into cell 6.  
Thanks to the very good coupling of HOMs to the cavity at room temperature, this method requires only a few watts of RF power flowing through the HOM cables. On the other hand, using a superposition of modes from the $\mathrm{TM_{010}}$ passband would have required an enormous amount of power to ignite the plasma, causing the risk of damaging and deteriorating the high-power antenna and its parts. This was not the case for the SNS cavities in which the fundamental power coupler (FPC) is set to match much heavier beam loading, therefore the FPC almost matches the Qo of the warm cavity making possible to ignite plasma using the $\mathrm{TM_{010}}$ passband modes.
\begin{figure}[t]
    \centering
    {\includegraphics[width=0.6\columnwidth]{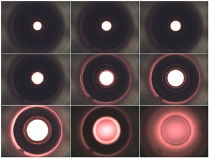}} \quad
	\caption{Ne plasma in each cell (9 to 1) of a LCLS-II cavity, upstream side view. \cite{berrutti2019plasma}}
	\label{fig:Plasma}
\end{figure}

For the first set of experiments of plasma ignition, a 1.3 GHz 9-cell cavity was equipped with view-ports on the two ends so that the plasma could be monitored through an optical camera and its location along the cavity easily confirmed. Picture of the plasma ignited inside each cell of a a 1.3 GHz 9-cell cavity is shown in Figure \ref{fig:Plasma}. 

However, to successfully clean SRF cavities in situ in cryomodules, it is necessary to detect in which cell the plasma glow discharge is located without looking at the camera image. For this purpose, a Labview program was developed to identify the plasma location by measuring the frequency shifts of the modes from the first dipole band (from 1-1 to 1-7) and comparing it with the simulated values. The simulated values are calculated by running finite element simulations in which a perturbation is introduced in each cavity cell by setting the real part of the dielectric constant to less than unity.

Plasma processing experiments focused on verifying the effectiveness of the methodology in abating field emission in SRF cavities, without affecting their performance. 
\begin{figure}
    \centering
    {\includegraphics[width=1\columnwidth]{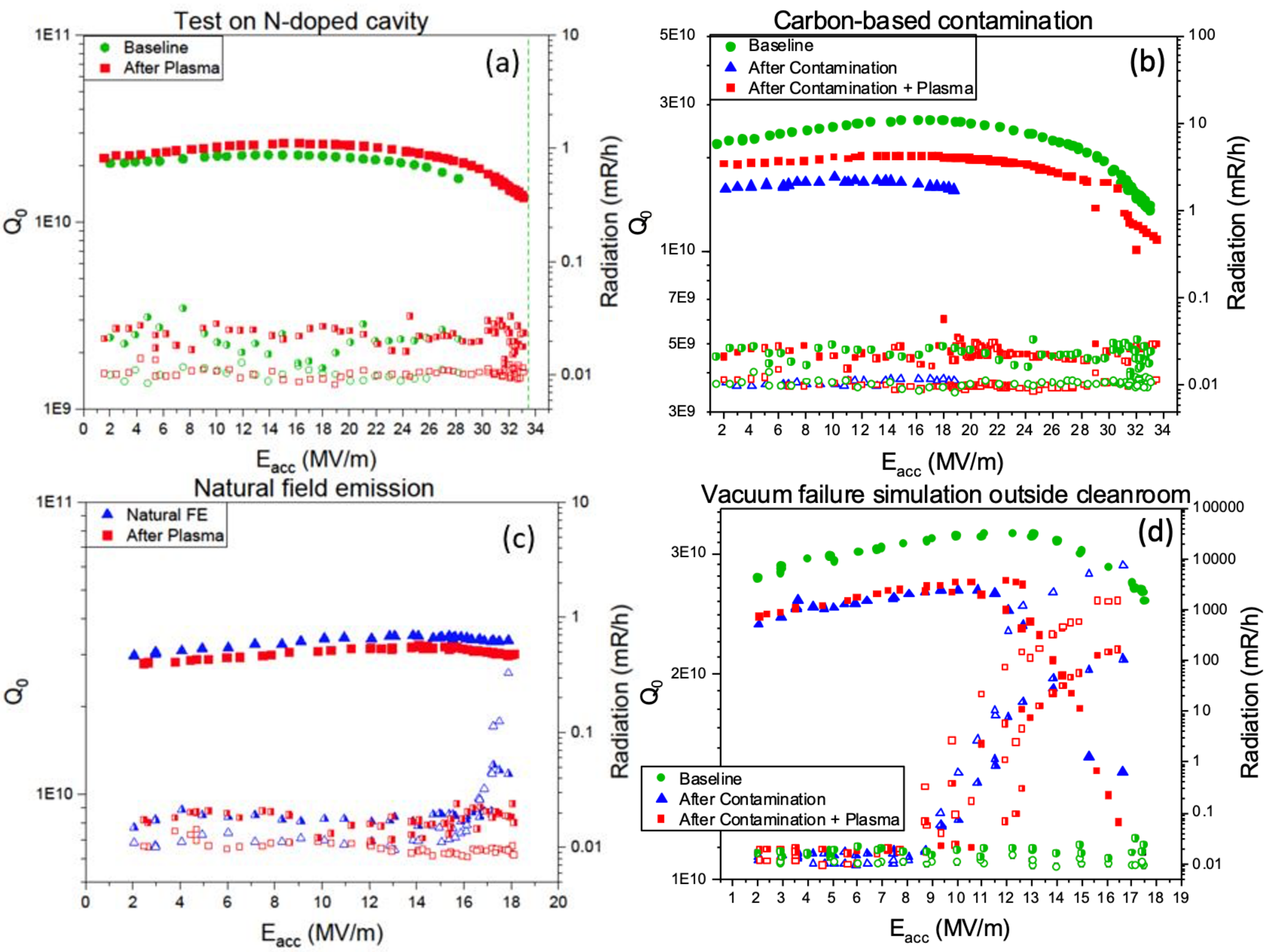}} \quad
	\caption{Examples of Q-factor ($Q_0$) versus accelerating field ($E_{acc}$) measurements performed at $T = 2$ K, before and after plasma processing, on: (a)  a N-doped single-cell cavity (during the baseline rf test the quench field was reached at $E_{acc}=33.5$ MV/m at 1.4 K, in agreement with the after plasma measurement. Q-factor values are also very similar indicating that the plasma cleaning did not affected the N-enriched layer.); (b) a carbon contaminated single-cell cavity; (c) a naturally field emitting 9-cell cavity; (d) a 9-cell cavity subjected to a simulated vacuum failure outside the clean-room. For all these graphs solid symbols are used for the Q0 value, empty symbols for the radiation level detected from the top detector, and half-filled symbols for the bottom radiation detector. \cite{giaccone2021field}}
	\label{fig:Results}
\end{figure}
The process was performed by injecting a mixture of neon and oxygen (few percentages) on one end of the cavity and pumping from the other end. The working pressure was usually maintained between 50 and 100 mTorr. 

Plasma cleaning was first applied to a N-doped cavity \cite{grassellino2013nitrogen}, that was RF tested before and after the processing to compare its performance (Figure \ref{fig:Results} (a)). The test demonstrated that the high Q-factors were fully preserved after the processing. Therefore, it was possible to conclude that the action of the plasma on the cavity surface does not affect the N-enriched layer typical of N-doped cavities.  
In addition, plasma cleaning was applied to multiple LCLS-II cavities with natural FE or artificially contaminated. The comparison between the RF tests conducted before and after plasma cleaning showed an increase in performance in the carbon contaminated single-cell (Figure \ref{fig:Results} (b)), in one out of two naturally field emitting cavities (Figure \ref{fig:Results} (c)) and in one nine-cell cavity exposed to vacuum failure simulation inside the cleanroom. A second cavity with natural FE was processed, but still showed x-ray activity after the plasma (Figure \ref{fig:Results} (d)), suggesting that the source of FE may not be C-related in this case, but due to metal flakes or surface defects. The three cavities used for vacuum failure simulation outside the cleanroom have shown little or no improvement attributable to plasma processing.  
From these studies it appeared that plasma processing was successful when applied in cavities with FE onset registered at high field levels (above 16 MV/m), while it was not successful when the onset started at low fields. This can be related with the source of FE: the plasma processing methodology applied in these experiments is effective against hydrocarbon contamination, but not metal flakes, which represent the most plausible cause of FE at low fields. Preliminary results from the analysis of particles collected from a single-cell cavity also suggested that metal flakes were introduced into the cavity during the vacuum failure experiment performed outside the cleanroom.

At Fermilab, some experiments of plasma ignitions on a PIP-II Single Spoke Resonator were also conducted and preliminary plasma ignition studies on a 112 MHz SRF Gun are on-going. The latter studies are conducted as a part of a collaboration with Brookhaven National Laboratory (BNL). 


\subsection{Plasma cleaning of SRF cryomodules}

Plasma processing was applied to the LCLS-II-HE verification cryomodule (vCM) and the successful results are discussed in Giaccone \textit{et al.} \cite{giaccone2022plasma}.
 
Before plasma processing the vCM underwent extensive tests, the results are reported in Posen \textit{et al.} \cite{posen2021lcls}. The vCM showed record performance in terms of both quality factor and accelerating gradient, exceeding the project specification. Particularly relevant for plasma processing, only one cavity (CAV5) exhibited detectable radiation ($0.6$mR/hr), but the FE source was processed during later testing, leaving the vCM field emission-free.
Nevertheless, the vCM represented a unique opportunity to scale the \SI{1.3}{\giga\hertz} plasma processing technique from a single TESLA-style \cite{aune2000superconducting} cavity to an entire cryomodule. For this reason, we plasma processed four out of the eight cavities of the vCM and we compared the RF cavity performance before and after plasma processing. 

\begin{figure}
    \centering
    {\includegraphics[width=1\columnwidth]{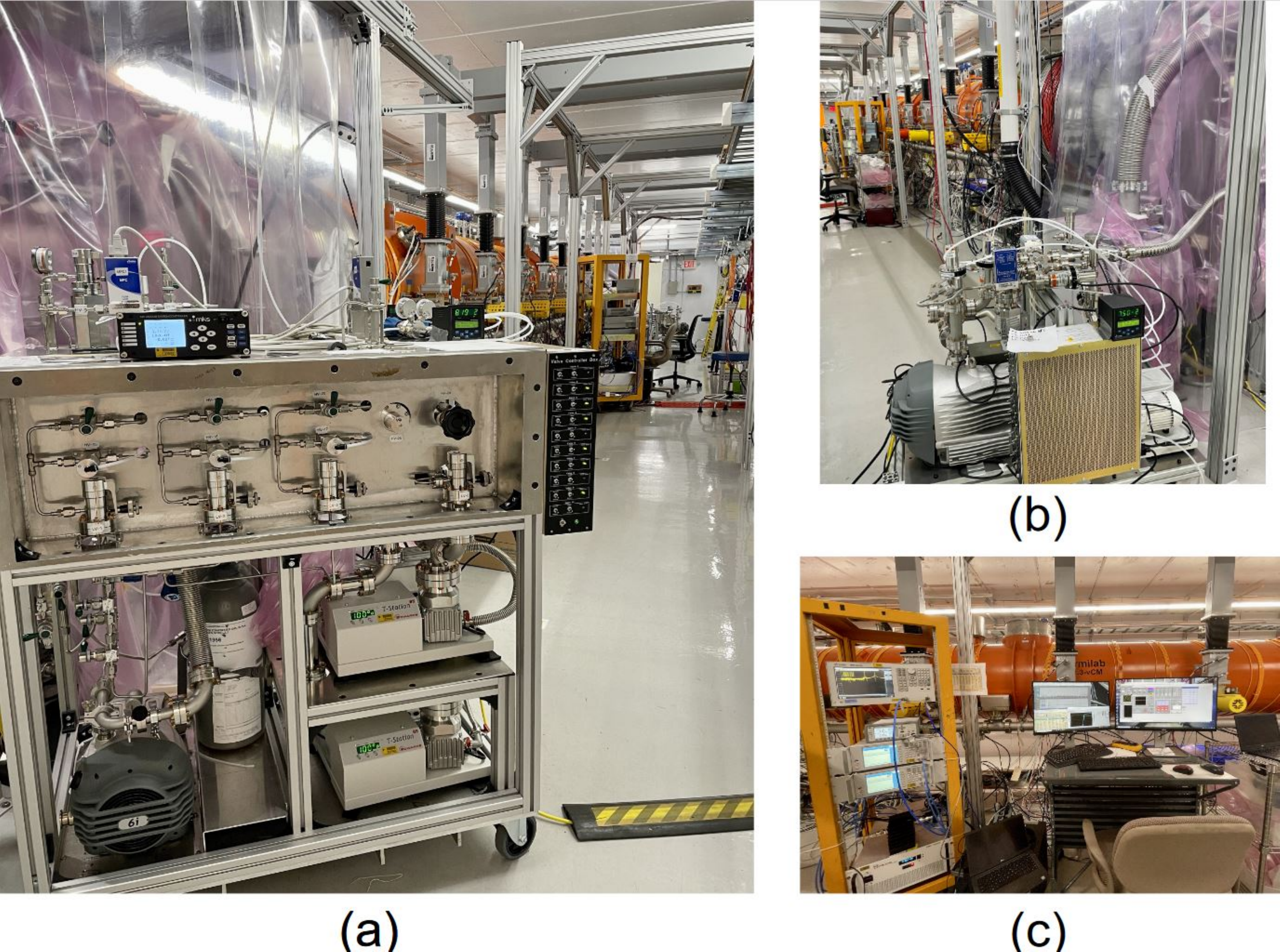}} \quad
	\caption{Experimental setup used to plasma clean cavities of the LCLS-II-HE vCM. The gas station is connected on one end of the CM (Fig. (a)), the vacuum and RGA station to the other end (Fig. (b)). The RF rack and computers are located in between the gas and vacuum stations (Fig. (c)). \cite{giaccone2022plasma}}
	\label{fig:SetUpCryo}
\end{figure}

Plasma cleaning of the vCM took place at the Cryomodule Test Facility (CMTF) at FNAL. Photos of the experimental system used to plasma process the vCM are shown in Fig. \ref{fig:SetUpCryo}. The gas injection system was connected to the upstream side of the vCM (near CAV1) and the vacuum cart on the downstream side (near CAV8). The connections between the gas and vacuum carts and the vCM were carried out in a class-100 portable cleanroom to minimize the risk of particle contamination. 
During plasma cleaning, a mixture of neon and oxygen in gaseous form is injected in the cavity string and a constant flow is maintained from CAV1 to CAV8. The glow discharge is ignited in one cell at a time following the direction of the gas flow; the cavities were also processed following the direction of the gas flow, from CAV1 to CAV8.

After plasma processing and evacuation, the vCM was cooled down to be RF tested again. The scope of this test was to verify that plasma processing did not cause any deterioration in cavity performance and to understand if it had an impact on MP. The plan was to remeasure the quality factor, the maximum and the usable gradient for each cavity, the presence of X-rays and dark current and the stability of operations at $20.8\,$MV/m. The time necessary to process multipacting was also investigated.

The results of the vCM prior to plasma processing are reported in table \ref{tab:vCMPerformance}, along with the comparison with the performance measured after plasma cleaning. For all eight cavities, it was verified that the performance was preserved both in terms of quality factor and accelerating gradient. This demonstrates that plasma processing did not introduce any observably detrimental contamination or particulates inside the cavity string, maintaining the vCM field emission-free. After plasma processing, the CM test confirmed the total module voltage of \SI{210}{\mega\volt}, versus the \SI{173}{\mega\volt} required by the LCLS-II-HE specification. The vCM average quality factor still exceeded the specification ($Q_0 = 2.7 \times 10^{10}$ at $20.8\,$MV/m), with an average of $Q_0 = 3.1 \times 10^{10}$. As explained in Posen \textit{et al.} \cite{posen2021lcls}, the administrative limit for the gradient in case the ultimate quench field of the cavity was not reached was set to $26.0\,$MV/m, while the maximum gradient defines the field level at which a cavity would  quench consistently without allowing any further field increase. The usable gradient instead is defined as the maximum field at which i) the cavity can operate for more than one hour without quenching, ii) the radiation stays below $50\,$mR/hr and iii) the cavity is 0.5 MV/m below its ultimate quench field limit. 

\begin{table*}
    \caption{\label{tab:vCMPerformance}Comparison of vCM performance measured before and after plasma processing. The four cavities to which plasma processing was applied are highlighted in bold.\cite{giaccone2022plasma}}
\small
\begin{tabulary}{\columnwidth}{c|CCCC|CCCC}
\hline \hline
&
\multicolumn{4}{c}{Before plasma} \vline &  \multicolumn{4}{c}{After plasma} \\
Cavity & $\mathrm{E_{acc}^{Max}}$ \small{(MV/m)} & $\mathrm{E_{acc}^{Usable}}$ \small{(MV/m)} & $\mathrm{Q_0}$ \small{at $21\,$MV/m} & MP & $\mathrm{E_{acc}^{Max}}$ \small{(MV/m)}  & $\mathrm{E_{acc}^{Usable}}$ \small{(MV/m)}  & $\mathrm{Q_0}$ \small{at $21\,$MV/m} & MP\\
\hline 
\textbf{1} & \textbf{23.4} & \textbf{22.9} & \textbf{3.0} & \textbf{Y} & \textbf{23.8} & \textbf{23.3} & \textbf{3.4} & \textbf{N} \\
2 & 24.8 & 24.3 & 3.0 & Y & 25.2 & 24.7 & 3.2 & Y \\
3 & 25.4 & 24.9 & 2.6 & Y & 26.0 & 26.0 & 3.4 & Y \\
\textbf{4} & \textbf{26.0} & \textbf{26.0} & \textbf{3.2} & \textbf{Y} & \textbf{26.0} & \textbf{26.0} & \textbf{3.2} & \textbf{N} \\
\textbf{5} & \textbf{25.3} & \textbf{24.8} & \textbf{2.9} & \textbf{Y} & \textbf{25.5} & \textbf{25.0} & \textbf{2.8} & \textbf{N} \\
6 & 26.0 & 25.5 & 3.4 & Y & 26.0 & 26.0 & 3.2 & Y \\
7 & 25.7 & 25.2 & 3.4 & Y & 25.9 & 25.4 & 3.3 & Y \\
\textbf{8} & \textbf{24.4} & \textbf{23.9} & \textbf{2.7} & \textbf{Y} & \textbf{24.7} & \textbf{24.2} & \textbf{2.6} & \textbf{N} \\ \hline
Average & 25.1 & 24.7 & 3.0 & & 25.3 & 25.1 & 3.1 & \\
Total & 209 & 205 & & & 210 & 208 & & \\
\hline \hline
\end{tabulary}
\end{table*}

As shown in Fig.$\,$(10) and (11) from Posen \textit{et al.} \cite{posen2021lcls}, during the first vCM RF test all the eight cavities suffered from multipacting induced quenches, during both the power rise to $21\,$MV/m (Fig. (10) \cite{posen2021lcls}) or the long duration operation at high gradient (Fig. (11) \cite{posen2021lcls}). As highlighted by the last column in table \ref{tab:vCMPerformance}, after plasma cleaning the four cavities that were not processed still showed MP quenches, while the four processed cavities did not experience any MP-related quench. 

These results demonstrate that plasma cleaning can fully eliminate multipacting in cavities in cryomodules. Therefore, performing plasma processing after a cryomodule has been assembled may decrease significantly its overall testing time and the subsequent accelerator commissioning time, reducing the overall project cost. In addition, cryomodules with multipacting-free cavities would have more stability during the accelerator operation, increasing its reliability.

\section {Plasma processing effort in the SRF community}
Many laboratories across the world are working to adapt plasma processing to SRF cavities with different geometries, different couplers and different resonant frequencies.

Having developed plasma processing for SNS cavities, Doleans \textit{et al.} applied the technique to ten cryomodules (eight of them high-beta CMs, two medium-beta CMs), either in offline facilities or directly in the linear accelerator (linac) tunnel. When applying plasma processing to the SRF cavities, they measured a peak in by-products partial pressure followed by a significant reduction over time, confirming the cleaning of the cavity surface from adsorbate gases. The plasma processed cavities showed in increase in accelerating gradient varying from 0.5 to 6.5MV/m, with an average $\mathrm{E_{acc}}$ increase of 2.4 MV/m. Worth to mention is that after three years of operation most of the processed cavities are still showing good performance.

Starting from 2019, Jefferson Laboratory (JLab) worked on developing plasma processing for CEBAF SRF cavities. The initial effort is focusing on C100 cryomodules, as reported in Powers \textit{et al.} \cite{powersplasma}. Following FNAL experience, JLab implemented a method to characterize and locate the plasma based on S21 VNA measurements. Once the glow discharge is ignited, the dielectric constant decreases and the change reflects in a frequency shift on each resonant mode proportional to the intensity of the field in the cell of ignition \cite{berrutti2019plasma}. JLab developed a software to measure a baseline S21 spectrum (with no plasma) and compare it with a live spectrum to compute the frequency shifts and identify the ignition cell of the plasma without a camera. In order to ignite the glow discharge inside the cavity RF volume, Powers \textit{et al.} use HOM belonging to the TE111 passband. Similarly to FNAL, they use higher order modes and couplers to ignite the glow discharge in the central cell and then move it to adjacent cells using a superposition of two modes. Powers \textit{et al.} applied plasma processing to multiple cavities, both individual cavities in a test-bench setup, both to cavities in the C100-10 CM. On individual cavities they saw a shift by almost 3MV/m in FE threshold, moving the operating gradient with 100 mRem/hr from 16.8 to 19.5MV/m. They also applied plasma processing to the C100-10 CM. Prior to plasma processing, this CM was removed from CEBAF and scheduled for disassembly since its performance was degraded due to catastrophic failures of two Viton O-rings, which caused cavities 1 and 2 to no longer being operated stably. Plasma processing was applied to all eight cavities, with an average increase in FE onset by 0.4MV/m and an increase by 0.3MV/m in 100mRem/hr FE levels. Just like the Fermilab vent to air test, this test proved that plasma processing can not be used to recover a cryomodule that has suffered severe damage. That being said, it was a useful exercise in order to validate the JLab plasma processing procedure and to train staff members. Jlab will be processing a C100 cryomodule in the spring of 2022 that has not suffered from a catastrophic failure.  Future work also includes a study of an ignition method to process C50 and C75 CEBAF cavities, effort on refining the software and the processing technique for C100 cryomodules, and on investigating novel plasma processing methods to mitigate non-hydrocarbon-related FE.

Wu \textit{et al.} \cite{wu2018situ, wu2019cryostat} developed plasma processing for the Chinese-Accelerator Driven System (C-ADS). C-ADS linac was successfully commissioned up to 25MeV, however field emission was a limiting factor. For this reason, both plasma and helium processing were explored as complementary in-situ techniques to mitigate FE. Huang \textit{et al.} conducted studies at the Institute of Modern Physics (IMP), Chinese Academy of Sciences on low-beta half wave resonators (HWRs) \cite{huang2019effect}. One the cryomodules (CM4) of the superconducting proton linac (CAFe) was accidentally vented after being subjected to helium processing. Although the performance degradation due to the accidental venting was not quantified, as it was not possible to test the CM again due to cryogenic system overhaul, they decided to apply plasma cleaning to remove possible contaminations from the inner surface of the five CM4 cavities.  The performance of three out of the five cavities after plasma processing matched the previous performance, in two of these cavities the performance after plasma cleaning actually increased. The other two cavities did not recover the initial performance. This suggests that plasma processing effectively removed FE induced by carbon contamination, however, as expected, was not effective on FE caused by dust or metal flakes. At IMP they are exploring the combined use of plasma cleaning and helium processing to achieve full recovery of the RF performance.

At the Facility for Rare Isotope Beams (FRIB), Michigan State University Zhang \textit{et al.} are working on developing plasma processing for the FRIB SRF linac \cite{zhangplasma}. A feasibility study for FRIB CMs indicates that plasma processing can be achieved with the RF couplers installed on the CMs cavities. The initial development started on 322MHz HWR with $\beta = 0.53$. They studied the ignition of the glow discharge in the cavity volume with pulsed RF power using different noble gases and pressure. Vertical tests measured on a quarter wave resonator (QWR) with $\beta = 0.085$ showed that at $~$10.5MV/m the measured radiation decreased from 1850mR/h to 4mR/h after plasma processing. They plan to demonstrate that the plasma ignition and the full procedure can be accomplished in FRIB cavities using the fundamental power coupler (which is weakly coupled).

University of Wisconsin also conducted an effort to apply plasma processing on their 200 MHz Quarter Wave Resonator (QWR) SRF electron gun, in 2012. They used a mixture of argon and oxygen and design a special antenna to optimize the coupling with the cavity at room temperature. Plasma processing significantly improved the field emission characteristics of the QWR SRF gun cavity, bringing their maximum accelerating field level approximately from 5 to 25 MV/m \cite{legg_plasma}. 

\section {Future perspective and conclusions}

The plasma ignition methodology is dependent on the cavity RF frequency, type of RF antennas and RF coupling. Because of this, a proper methodology shall be developed for each type of cavity and cryomodule to be plasma processed. 

The current plasma processing is based on the cleaning of the SRF inner surface from carbon contamination via a mixture of neon and oxygen plasma. Therefore, in order to reach the maximum accelerating gradient of a cavity without field emission, the processing needs to be extended to the use of different and more aggressive gas mixtures to be able to clean the inner cavity surface from a large variety of contaminants.

Summarizing, plasma processing may be useful to abate field emission potentially to all types of SRF cavities, enabling the reach of higher energies on present and future machines. In addition, cavities characterized by strong multipacting bands that usually require several hours to be processed in cryomodule, will strongly benefit from plasma cleaning. It was already demonstrated that plasma cleaning can completely eliminate MP in cryomodule in TESLA-type cavities, therefore it could potentially save all the effort usually needed to process MP in these structures. 

An aggressive R\&D effort should then been pursued in the SRF community with the goal of: extending the applicability of plasma cleaning to a large number of cavity and cryomodule types, increasing the processing effectiveness against a large variety of field emitters, and optimizing the processing effectiveness against multipacting.

\section {Acknowledgements}

Work supported by the Fermi National Accelerator Laboratory, managed and operated by Fermi Research Alliance, LLC under Contract No. DE-AC02-07CH11359 with the U.S. Department of Energy.




\bibliographystyle{JHEP}
\bibliography{myreferences}  




%

\end{document}